\documentstyle[natbib,epsf,psfrag,ph]{mn}
\newcommand{\thz}{$\Delta\theta$-$z_{\rm s}$ relation}
\psfrag{Zs}{\large $z_{\rm s}$}
\psfrag{DJ}{\large $\Delta\theta$}
\title
[
The \thz\ as a cosmological test 
]
{
The \thz\ for gravitational lenses as a cosmological test 
}

\author
[
Phillip Helbig
]
{
Phillip Helbig\thanks{email: {\tt p.helbig@jb.man.ac.uk}}
\\
University of Manchester, Nuffield Radio Astronomy Laboratories, 
Jodrell Bank, Macclesfield, Cheshire SK11~9DL, UK
}

\date{
Accepted.  Received 
}

\pagerange{\pageref{firstpage}--\pageref{lastpage}}

\pubyear{1998}

\begin{document}\label{firstpage}
\maketitle

\begin{abstract}
Recently, \citet{MParkJGott97a} claimed that there is a statistically
significant, strong, negative correlation between the image separation
$\Delta\theta$ and source redshift $\zs$ for gravitational lenses.  This
is somewhat puzzling if one believes in a flat ($k=0$) universe, since
in this case the typical image separation is expected to be independent
of the source redshift, while one expects a negative correlation in a
$k=-1$ universe and a positive one in a $k=+1$ universe. 
\citeauthor{MParkJGott97a} explored several effects which could cause
the observed correlation, but no combination of these can explain the
observations with a realistic scenario.  Here, I explore this test
further in three ways.  First, I show that in an inhomogeneous universe
a negative correlation is expected regardless of the value of $k$.
Second, I test whether the \thz\ can be used as a test to determine
$\lnull$ and $\onull$, rather than just the sign of $k$.  Third, I
compare the results of the test from the \citeauthor{MParkJGott97a}
sample to those using other samples of gravitational lenses, which can
illuminate (unknown) selection effects and probe the usefulness of the
\thz\ as a cosmological test. 
\end{abstract}


\section{Introduction}

Historically, there has been little interest in the \thz\ compared to
other cosmological tests based on gravitational lensing statistics,
perhaps because the inflationary paradigm \citep[\eg][]{AGuth81a}, which
began about the same time as the discovery of the first gravitational
lens \citep*{DWalshCW79a}, has become so influential.  Since a flat
($k=0$) universe is a robust prediction of inflation, many researchers
assume this and consider only flat universes (or, at most, $k=-1$
cosmological models with $\lnull=0$).  Due to the fact that for the
popular singular isothermal sphere model for a single-galaxy lens the
average 
image separation $\Delta\theta$,
integrated over the lens redshift $\zd$ from $\zd=0$ to $\zd=\zs$,
is {\em completely independent\/} of the
source redshift $\zs$ in a flat universe, there is little point in
pursuing the \thz\ if one is interested primarily in flat cosmological
models.  If one is not committed to a flat universe, then of course one
should not assume $k=0$, but even if one believes that the universe must
be flat, it is still important to test this belief observationally.  The
situation is somewhat worsened by the fact that most `standard'
cosmological tests such as the \mbox{$m$-$z$} 
(magnitude-redshift or `standard candle') and 
$\theta$-$z$ (angular size-redshift or `standard rod') relations, `conventional'
gravitational lensing statistics, age of the universe) are relatively
insensitive to the radius of curvature of the universe ($R_{0} \sim
(|\Omega_{0}+\lambda_{0}-1|)^{-\frac{1}{2}}$), being degenerate in
combinations of $\lnull$ and $\onull$ in directions roughly
perpendicular to lines of constant $R_{0}$ in the $\lnull$-$\onull$
plane.  A notable exception are constraints derived from CMB
anisotropies \citep*[\eg][]{DScottSW95a,WHuSS97a}.

\section{Theory}

For a singular isothermal sphere lens, the angular image separation is
given by \citep*[\eg][]{ETurnerOG84a} 
\begin{equation}\label{tog-g}
\Delta\theta = 8\pi\left(\frac{v}{c}\right)^{2}\frac{\dds}{\ds} \quad,
\end{equation}
where $v$ is the velocity dispersion and $D$ is the angular size
distance (see below).  Even if the singular isothermal sphere is not a
perfect model for the gravitational lens systems considered, it is still
a good approximation when one is concerned only with the image
separation.  For a given $v$, by combining Eqs.~(5) and (6) in
\citet*{JGottPL89a} and using the more appropriate and more general
angular size distances, one obtains 
an expression for the average image separation $\Delta\theta$,
by integrating over the lens redshift $\zd$ from $\zd=0$ to $\zd=\zs$,
\begin{eqnarray}\label{big-g}
& & 
\left(\int\limits_{0}^{\zs}
\D\zd
\frac{
\dds^{3}\dd^{2}\left(1+\zd\right)^{2}
}
{
\ds^{3}Q
}
\right)
\nonumber \\
\frac{\Delta\theta(\zs)}{\Delta\theta(0)}&=&\rule[0.5ex]{4cm}{0.1mm}\quad,
\\
& & 
\left(\int\limits_{0}^{\zs}
\D\zd
\frac{
\dds^{2}\dd^{2}\left(1+\zd\right)^{2}
}
{
\ds^{2}Q
}
\right)\nonumber
\end{eqnarray} 
where 
\begin{equation}\label{little-g}
Q = \sqrt{\onull\left(1+\zd\right)^{3}-
\left(\onull+\lnull-1\right)\left(1+\zd\right)^{2}+\lnull} \quad.
\end{equation}
The $D_{ij}$ (with $D_{k}:=D_{0k}$) in Eqs.~(\ref{tog-g}) and
(\ref{big-g}) are angular size distances, which are functions of the
lens and source redshifts $\zd$ and $\zs$, the cosmological parameters
$\lnull$ and $\onull$ as well as the `homogeneity parameter' $\eta$,
which gives the fraction of smoothly, as opposed to clumpily,
distributed matter along the line of sight.  Note that Eq.~(\ref{big-g})
is valid for all combinations of $\lnull$, $\onull$ and $\eta$.  The
angular size distances can be computed for arbitrary combinations of
these parameters by the method outlined in \citet*{RKayserhs97a}. 

\begin{figure}
\epsfxsize=\columnwidth
\epsfbox{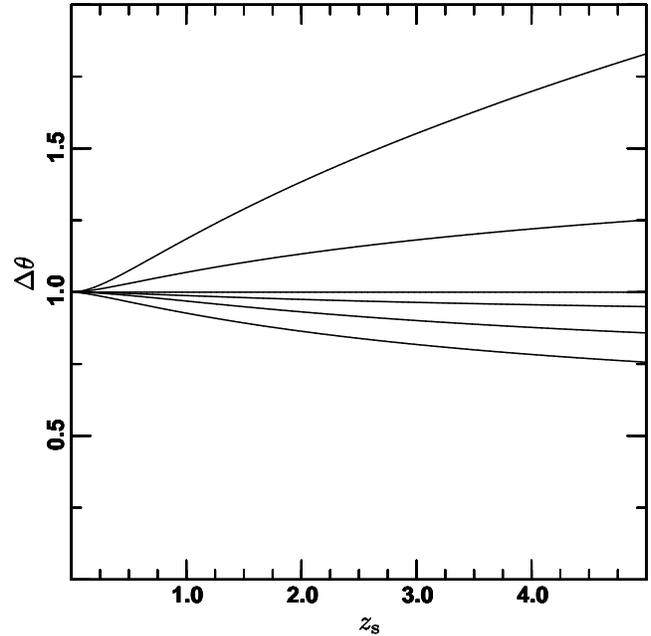}
\caption{Normalized image separation as a function of source redshift. 
From the top, the ($\lnull$,$\onull$) values are (2,4), (0,4), $k=0$,
(0,0.7), (0,0.3) and (-5,1).  For $k=0$ the result is valid for all
($\lnull$,$\onull$) values whose sum is 1.  $\eta=1$.  \label{eta1-f}} 
\end{figure}

\begin{figure} 
\epsfxsize=\columnwidth
\epsfbox{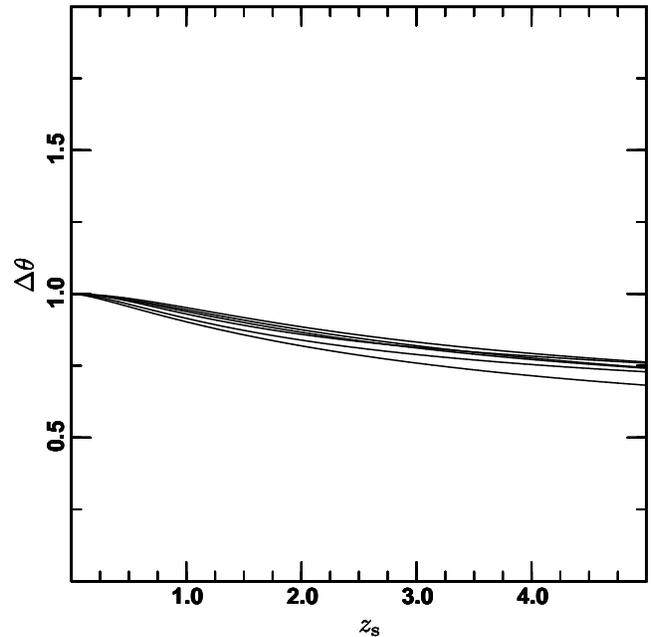}
\caption{The same as Fig.~\ref{eta1-f} except that here $\eta=0$.} 
\label{eta0-f}
\end{figure}

Figures \ref{eta1-f} and \ref{eta0-f} show $\Delta\theta$ as a function
of $\zs$ for various cosmological models, for $\eta=1$ (the traditional
case assuming a completely homogeneous universe) and $\eta=0$ as extreme
cases.  Note in Fig.~\ref{eta1-f} that the curve is a horizontal line
for $k=0$, has positive slope for $k=+1$ and negative slope for $k=-1$,
where $k:=\sign(\onull+\lnull-1)$.  In Fig.~\ref{eta0-f}, for $\eta=0$,
the slope is negative regardless of the value of $k$. 
Thus, at first sight it appears that an inhomogeneous universe, a 
possibility not
investigated by \citet[hereafter PG]{MParkJGott97a}, might be able to explain the
puzzling negative correlation between $\Delta\theta$ and $\zs$. 
However, it is shown in Sect.~\ref{rd} that even the extreme 
$\eta=0$ scenario produces an anticorrelation which is much weaker than 
that found by PG.
This effect can be qualitatively understood by realizing how
Eq.~(\ref{big-g}) is affected by decreasing $\eta$: inspection shows
that this might be estimated by examining $\dds/\ds$.  All other things
being equal, the angular size distance increases with decreasing $\eta$. 
Also, the effect of $\eta$ is more noticeable at large redshift
differences.  Since $\zs \geq \zs-\zd$, the denominator is the more 
important term, and so 
decreasing $\eta$ increases $\ds$ and so decreases $\dds/\ds$ and
thus $\Delta\theta(\zs)/\Delta\theta(0)$. 

\section{Data}

PG used an inhomogeneous sample of gravitational lenses from the
literature.  While this seems problematic at first sight, PG noted that 
there is no reason to believe that this should influence the analysis.  
Nevertheless, it is worth comparing the PG results to those obtained 
from a better defined sample.

The observational data provided by the JVAS and CLASS surveys offer an
independent sample of gravitational lenses.  JVAS is the Jodrell Bank
VLA Astrometric Survey \citep{APatnaikBWW92a}; CLASS is the Cosmic Lens
All-Sky Survey \citep{SMyersetal98a}.  Even though the observational
tasks are not yet complete, the JVAS and CLASS surveys which constitute
the database have already yielded sufficient gravitational lenses to enable
one to make an independent analysis.  Table~\ref{table-t} shows the
current state of knowledge about the JVAS/CLASS gravitational lenses. 
Note that the questionable source redshift for $2114+022$ is probably
the redshift of an additional lensing galaxy (this interpretation is
supported by several independent lines of evidence). 
\begin{table*}
\caption{The JVAS/CLASS gravitational lenses.}
\label{table-t}
\begin{tabular}{lcrcc@{\extracolsep{8pt}}l@{\extracolsep{0pt}}ccclcl}
\hline
\multicolumn{1}{c}{\rule[-2mm]{0mm}{6mm}Name} & 
\multicolumn{3}{c}{\# images} & 
\multicolumn{2}{c}{$\Delta\theta$} & & 
lens galaxy type & &
\multicolumn{1}{c}{$\zd$} & &
\multicolumn{1}{c}{$\zs$} \\ \nonumber
& & & & \multicolumn{2}{c}{[\arcsecond]} & & & & & \\
\hline
0218+357  && ring + 2 &&& 0.33 && spiral      && 0.6847 && 0.96 \\
0414+0534 && 4        &&& 2.0  && elliptical  && ?      && 2.62 \\
0712+472  && 4        &&& 1.2  && ?           && 0.406  && 1.339 \\
1030+074  && 2        &&& 1.6  && peculiar    && 0.599  && 1.535 \\
1422+231  && 4        &&& 1.2  && ?           && 0.65   && 3.62 \\
1600+434  && 2        &&& 1.4  && spiral      && 0.4144 && 1.589 \\
1608+656  && 4        &&& 2.2  && spiral?     && 0.64   && 1.39 \\
1933+503  && 4+4+2    &&& 0.9  && ?           && 0.755  && ?    \\
1938+666  && 4+2      &&& 0.9  && ?           && ?      && ?    \\
2045+265  && 4+1?     &&& 2.0  && ?           && 0.87   && 1.28 \\
2114+022  && 2+2?     &&& 2.4  && ?           && 0.316  && 0.588? \\
\hline
\end{tabular}
\end{table*}

\begin{figure}
\epsfxsize=\columnwidth
\epsfbox{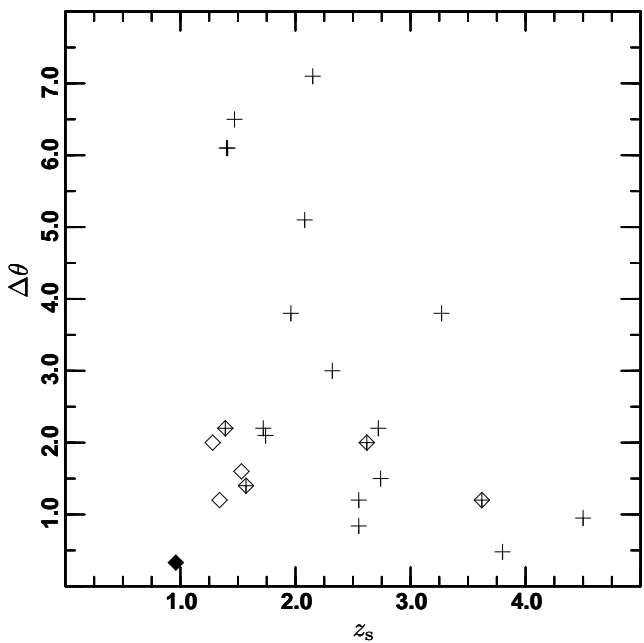}
\caption{Source redshifts $\zs$ and image separations $\Delta\theta$ (in 
arcsec) for 
the gravitational lens systems studied in this paper.  Crosses represent 
the PG sample (20 systems; note that two data points with
$\Delta\theta\approx 6$~arcsec almost coincide); 
diamonds represent the JVAS/CLASS sample (8 systems; of course only 
those with known source redshifts are included).
Note that there is an overlap of four data points.  The filled diamond 
represents the system $0218+357$, which was not used by PG although its 
source redshift had been published before the PG analysis was done 
\citep{CLawrence96a}.} 
\label{data-f}
\end{figure}

Although not all source redshifts in the JVAS/CLASS sample are known, 
8 out of 11 are, and based on our survey, discovery and followup 
strategies there is no reason to suspect the unknown source 
redshifts to be statistically different from those already known.
Figure \ref{data-f} shows the source redshifts and image separations of 
the gravitational lens systems used in this paper: the PG sample and the 
JVAS/CLASS sample.

\section{Calculations}

All calculations here implement the method of
PG, which uses the Spearman rank correlation
test to generate a relative probability for a given cosmological model. 
PG noted the fact that they always obtained a
low probability with their sample, even when allowing for non-flat
cosmological models (albeit in a limited area of parameter space),
galaxy evolution or departure from the singular isothermal sphere model.
As PG noted, allowing for these effects
increases the probability, since they all tend to create a negative
correlation in a flat universe, but the magnitude of the effect is not
large enough to explain the observations.  Again as noted by
PG, if the lenses are parts of clusters, then
this will work in the opposite direction, making the observed negative
correlation even more puzzling.

Calculations were done for four samples:
\begin{description}
\item the PG sample
\item the PG sample with the addition of the system \mbox{$0218+357$}
\item the JVAS/CLASS sample
\item the union of all samples
\end{description}
Note that the source redshift for $0218+357$ had been published before 
the PG analysis was done \citep{CLawrence96a}.  Since $0218+357$ lies 
below and to the left of all other data points, it is clear that 
including it will weaken the puzzling negative correlation found by
PG; this is discussed more quantitatively in Sect.~\ref{rd}.

\section{Results and discussion}
\label{rd}

Since the PG test assigns a low probability to a
$k=0$ universe, the question arises as to whether it can be used as a
general cosmological test to determine the values of $\lnull$ and
$\onull$.  This is not the case.
For all four samples I have calculated the Spearman rank correlation
probability as a function of $\lnull$ and $\onull$ in a range of
parameter space ($-8<\lnull<2$ and $0<\onull<10$) much larger than that
allowed even by a generous interpretation of observations.  This was
done with a resolution of $0.1$ in both $\lnull$ and $\onull$ for both 
$\eta=1$ and $\eta=0$. 
The
Spearman rank correlation probability is essentially constant over a
wide range of parameter space; basically, either all cosmological models
are probable, or all are improbable, depending on the sample used. 

The probability is a weak function of the cosmological model, with the 
sharpest transition occuring when crossing the $k=0$ line in the 
$\lnull$-$\onull$ plane.  For all samples except the 
PG sample, the probability is $>$$5\%$ in almost 
the entire parameter space;\footnote{For the JVAS/CLASS sample, the 
maximal probability is 0.955 and is realized in almost the entire $k=+1$ 
area of the parameter space.} 
those cosmological models with a lower 
probability are among those ruled out by current observations.  Thus, 
the Spearman 
rank correlation probability does not allow one to reject any otherwise 
viable cosmological models, which shows both that there is no reason to 
expect unknown effects in the gravitational lens samples and that it is 
not very useful as a cosmological test.  For the PG
sample, the 1\% contour corresponds almost exactly to the $k=0$ line, 
with higher values for a negatively curved universe.  Thus, the 
PG sample is {\em marginally\/} compatible with
a $k=-1$ cosmological model, although the probability values are low 
throughout the $\lnull$-$\onull$ plane, with values near the maximum 
of $0.025$ being attained only for small (but realistic) $\onull$ values 
and large (in absolute value) negative values of $\lnull$.
Since there are no known selection effects which can account for the
differences 
between the PG sample and other samples, 
either the test is not very useful and/or it is pointing to
unknown selection effects in the literature sample used by
PG.  The fact that the PG result changes dramatically (probability 
$\approx$ 10--20\% in most of the $\lnull$-$\onull$ plane) by the inclusion 
of just 1 additional data point, which could have been included in their 
analysis, argues in favour of the former possibility.

The above discussion was for $\eta=1$.  For $\eta=0$ the situation is
qualitatively the same and quantitatively involves only slightly
different values of probabilities derived from the Spearman rank
correlation test. 

It is interesting to compare the probabilities from the Spearman rank
correlation test for the PG sample using the
actual values of $\zs$ and $\Delta\theta$ as used by
PG to those obtained using more up-to-date data
for the {\em same\/} lens systems.  If two values are very near each
other, rounding them off to the same values produces a different result
for the rank correlation test than if they differ by even a small
amount.  Using more up-to-date data, an even lower probability is
obtained for the PG sample, for $\eta=1$ and
$\eta=0$, for a wide variety of cosmological models.

\section{Conclusions}

\citet{MParkJGott97a} pointed out that the image separations in
gravitational lens systems show a strong significant negative
correlation with the source redshift, while in a flat universe one would
expect no correlation (while a negative correlation would be expected in
a universe with negative curvature and a positive one in a universe of
positive curvature).  None of the possibilities they examined were
strong enough to explain the effect.  A possibility not examined by
them, namely an inhomogeneous universe, produces a negative correlation
regardless of the sign of the curvature, but it too is not strong enough
to account for the effect.  As a general test for the values of $\lnull$
and $\onull$ the test is of no use, all cosmological models being
assigned roughly the same probability, but {\em which\/} value they are
assigned depends on the sample used. 

The strong dependence of the result on the sample used seems to indicate
that the result of \citet{MParkJGott97a} is due not to some
physical cause but rather to unidentified selection effects in the
sample of gravitational lenses taken from the literature.  The large
number of JVAS and CLASS lenses gives us an independent comparison
sample, thus demonstrating the need for discovering a large number of
lenses in a well-defined sample.  As \citet{MParkJGott97a} point
out, since many conclusions based on `conventional' gravitational
lensing statistics are based on essentially the same lenses as in their
literature sample, if this sample is for some unknown reason atypical,
then conclusions drawn from statistical analyses of it must be examined
with care.  It will thus be interesting to see what conclusions can be
drawn from a statistical analysis of the JVAS/CLASS sample after the
observational tasks have been completed.  (We expect to find more
lenses, but have no qualms about using the present incomplete sample in
this analysis since there is no reason to believe that a larger sample
would show a different \thz.)

\section{Note}

Since this work was completed, two other responses to
\citet{MParkJGott97a} (apart from \citet{PHelbig98pressa}) have appeared. 
The first \citep{LWilliams97a} is complementary to this work in that it
assumes the effect is real and explores the astrophysical consequences
while the second \citep{ACooray98a} is more similar to this analysis,
arriving at essentially the same conclusions
though using different observational data (and
exploring neither the question of usefulness as a general test for
$\lnull$ and $\onull$ nor the effects of a locally inhomogeneous
universe).

\section{Acknowledgements}

I thank Asantha Cooray for comments on the manuscript and my 
collaborators in the CERES project for helpful discussions.  This research was
supported by the European Commission, TMR Programme, Research Network
Contract ERBFMRXCT96-0034 `CERES'.

\bibliographystyle{mnras}
\bibliography{abbrev,astro,astro_ph}

\begin{thebibliography}{14}
\expandafter\ifx\csname natexlab\endcsname\relax\def\natexlab#1{#1}\fi

\bibitem[{Cooray(1998)}]{ACooray98a}
Cooray A., 1998, AJ, submitted (astro-ph/9711179)

\bibitem[{Gott et~al.(1989)Gott, Park \& Lee}]{JGottPL89a}
Gott III J.R., Park M.G., Lee H.M., 1989, ApJ, 338, 1

\bibitem[{Guth(1981)}]{AGuth81a}
Guth A.H., 1981, Phys. Rev. D, 23, 347

\bibitem[{Helbig(1998)}]{PHelbig98pressa}
Helbig P., 1998, In M{\"u}ller V., ed., Large Scale Structure: Tracks and
  Traces, World Scientific, Singapore, in press

\bibitem[{Hu et~al.(1997)Hu, Sugiyama \& Silk}]{WHuSS97a}
Hu W., Sugiyama N., Silk J., 1997, Nat, 386, 37

\bibitem[{Kayser et~al.(1997)Kayser, Helbig \& Schramm}]{RKayserhs97a}
Kayser R., Helbig P., Schramm T., 1997, A\&A, 318, 680

\bibitem[{Lawrence(1996)}]{CLawrence96a}
Lawrence C.R., 1996, In Kochanek C.S., Hewitt J.N., eds., Astrophysical
  Applications of Gravitational Lensing, Kluwer Academic Publishers, Dordrecht,
  p. 299

\bibitem[{Myers et~al.(1998)}]{SMyersetal98a}
Myers S.T., et~al., 1998, in preparation

\bibitem[{Park \& Gott(1997)}]{MParkJGott97a}
Park M.G., Gott III J.R., 1997, ApJ, 489, 476 (PG)

\bibitem[{Patnaik et~al.(1992)Patnaik, Browne, Wilkinson \&
  Wrobel}]{APatnaikBWW92a}
Patnaik A.R., Browne I.W.A., Wilkinson P.N., Wrobel J.M., 1992, MNRAS, 254, 655

\bibitem[{Scott et~al.(1995)Scott, White \& Silk}]{DScottSW95a}
Scott D., White M., Silk J., 1995, Sci, 268, 829

\bibitem[{Turner et~al.(1984)Turner, Ostriker \& Gott}]{ETurnerOG84a}
Turner E.L., Ostriker J.P., Gott III J.R., 1984, ApJ, 284, 1

\bibitem[{Walsh et~al.(1979)Walsh, Carswell \& Weymann}]{DWalshCW79a}
Walsh D., Carswell R.F., Weymann R.J., 1979, Nat, 279, 381

\bibitem[{Williams(1997)}]{LWilliams97a}
Williams L.L.R., 1997, MNRAS, 292, L27

\end{thebibliography}

\bsp
\label{lastpage}
\end{document}